\documentclass{article}
\usepackage{txfonts}
\usepackage{color}
\usepackage{comment}
\usepackage{cmll}
\usepackage{amsfonts}
\usepackage{theorem}
\usepackage{graphicx}
\usepackage{subfigure}
\usepackage{QED}
\usepackage{diagrams}

\newcommand{\modif}[1]{#1}

\newarrow{Corresponds}<--->
\newtheorem{definition}{Definition}
\newtheorem{theorem}{Theorem}
\newtheorem{fact}{Fact}
\newtheorem{lemma}{Lemma}
\newtheorem{corollary}{Corollary}

\newcommand{\paths}[1]{\mathcal{P}(#1)}
\newcommand{\pathsports}[3]{\mathcal{P}_{#2 \rightarrow #3}(#1)}
\newcommand{\pathsfree}[1]{\mathcal{P}_f(#1)}
\newcommand{\pathslong}[2]{\mathcal{P}_{#1}(#2)}

\newcommand{\crossings}[2]{\chi_{#1}(#2)}
\newcommand{\pathred}[2]{\delta_{#1}(#2)}
\newcommand{\EX}[1]{\mathsf{EX}(#1)}
\newcommand{\NEX}[2]{\mathsf{NEX}_{#1}(#2)}
\newcommand{\idem}[1]{\langle #1 \rangle}

\newcommand{\orth}[0]{\perp}
\newcommand{\dorth}[0]{\overline{\perp}}
\newcommand{\comm}[0]{\leftrightarrow}
\newcommand{\comms}[0]{
	\stackrel{\star}{\leftrightarrow}}

\newcommand{\names}[0]{\mathcal{N}}
\newcommand{\namesr}[1]{\names (#1)}
\newcommand{\tokens}[0]{\mathfrak{T}}

\newcommand{\lstar}[0]{{L^\star}}
\newcommand{\dlstar}[0]{\partial \lstar}
\newcommand{\dlstarpme}[0]{{\dlstar}^+_{me}}
\newcommand{\dlstarpa}[0]{{\dlstar}^+_{a}}

\newcommand{\rop}[1]{r_{#1}}
\newcommand{\sop}[1]{s_{#1}}
\newcommand{\dop}[2]{d_{#1,#2}}
\newcommand{\uop}[1]{u_{#1}}
\newcommand{\vop}[1]{v_{#1}}
\newcommand{\eop}[1]{e_{#1}}
\newcommand{\pop}[0]{p}
\newcommand{\qop}[0]{q}
\newcommand{\starred}[1]{{#1}^\star}
\newcommand{\rsop}[1]{\starred{\rop{#1}}}
\newcommand{\ssop}[1]{\starred{\sop{#1}}}
\newcommand{\dsop}[2]{\starred{\dop{#1}{#2}}}
\newcommand{\usop}[1]{\starred{\uop{#1}}}
\newcommand{\vsop}[1]{\starred{\vop{#1}}}

\newcommand{\Mlstar}[0]{\mathfrak{L}}
\newcommand{\Mdlstar}[0]{\mathfrak{d} \Mlstar}

\newcommand{\Mrop}[1]{\mathfrak{r}_{#1}}
\newcommand{\Msop}[1]{\mathfrak{s}_{#1}}
\newcommand{\Mdop}[2]{\mathfrak{d}_{#1,#2}}
\newcommand{\Muop}[1]{\mathfrak{u}_{#1}}
\newcommand{\Mvop}[1]{\mathfrak{v}_{#1}}
\newcommand{\Meop}[1]{\mathfrak{e}_{#1}}
\newcommand{\Mpop}[0]{\mathfrak{p}}
\newcommand{\Mqop}[0]{\mathfrak{q}}

\newcommand{\Mrsop}[1]{\starred{\Mrop{#1}}}

\newcommand{\Mdsop}[2]{\starred{\Mdop{#1}{#2}}}
\newcommand{\Musop}[1]{\starred{\Muop{#1}}}

\newcommand{\ms}[0]{M}
\newcommand{\es}[1]{E_{#1}}
\newcommand{\ewns}[0]{\es{\wn}}
\newcommand{\eocs}[0]{\es{\oc}}
\newcommand{\idxf}[0]{f}

\newcommand{\token}[4]{(#1, #2, #3, #4)}
\newcommand{\dtoken}[0]{\token{\ms}{\ewns}{\eocs}{\idxf}}

\newcommand{\funextends}[3]{#1[#2 \mapsto #3]}

\newcommand{\concat}[0]{\bullet}

\newcommand{\bzero}[0]{0}
\newcommand{\bone}[0]{1}
\newcommand{\bool}[0]{\{\bzero,\bone\}}

\newcommand{\words}[1]{{#1}^\star}

\newcommand{\finfuncset}[2]{\mathcal{F}_{fin}(#1, #2)}
\newcommand{\operation}[1]{
\stackrel{#1}{\longrightarrow}
	}

\begin{document}

\title{The Geometry of Interaction of Differential Interaction Nets}
\author{Marc de Falco
\footnote{Work supported by the French ANR project ``Choco''.} \\
Institut de Math\'ematiques de Luminy\\
Universit\'e d'Aix-Marseille\\
\textsf{defalco@iml.univ-mrs.fr}
}

\maketitle
\thispagestyle{empty}

\begin{abstract}
    The Geometry of Interaction purpose is to give a semantic of
    proofs or programs accounting for their dynamics. The initial
    presentation, translated as an algebraic weighting of paths in
    proofnets, led to a better characterization of the 
    $\lambda$-calculus optimal reduction. 
    Recently Ehrhard and Regnier
    have introduced an extension of the Multiplicative Exponential fragment
    of Linear Logic (MELL) that is able to express
    non-deterministic behaviour of programs and a proofnet-like
    calculus: Differential Interaction Nets.
    This paper constructs a proper
    Geometry of Interaction (GoI) for this extension.
    We consider it both as an algebraic theory and as a concrete 
    reversible computation.
    We draw links between this GoI and the one of MELL.
    As a by-product we give for the first time
    an equational theory suitable for the GoI of the Multiplicative
    Additive fragment of Linear Logic.
\end{abstract}

\tableofcontents

\section{Introduction}

The purpose of this paper is to extend the
\emph{Geometry of Interaction} (GoI), a special kind of semantics,
to a paradigm of non-deterministic
computations: the \emph{differential interaction nets}.

The Geometry of Interaction was introduced by Girard in \cite{Girard89}.
The author
presented it as a new kind of semantics for linear logic accounting for
the dynamics of cut-elimination. Although the original presentation
was in terms of bounded operators over a Hilbert space, it could be
reformulated as a characterization of a certain class of paths in a graph-like
structure associated to programs/proofs allowing the interpretation of
these programs/proofs as reversible automata \cite{DanosRegnier96}.
The idea of 
studying paths in a proof or in a program is natural, be it to express
syntactic properties, like an access path to variable definition, or 
dynamic properties, like a path unfolding all abstractions over a subroutine
call. For such path to be relevant it needs to be well-behaved with respect
to the execution, that is to say it needs to be somehow invariant.
Invariance has a big computing cost because it needs to fully compute
the execution to be able to extract \emph{a posteriori} this information.
The GoI solves this issue by allowing to compute \emph{a priori} those paths.
Moreover it gives an algebraic translation of this problem, and allows
to assert that there is essentially one good notion of paths, as 
can be seen in \cite{AspertiDanosLaneveRegnier94}.

Since its introduction as a semantics of the Multiplicative Exponential fragment
of Linear Logic (MELL), the GoI has been extended to the additives giving
a semantics of whole linear logic, see \cite{Girard95}. To define it,
the author has used a notion of isomorphic semantics: \emph{variance}, and
given definitions and properties up to it. In this process, it seems
that the nice property of being an algebraic characterization of relevant
paths in programs was lost. Although a model as reversible computations
was extracted from this presentation in \cite{Laurent01c}, an equational
theory for the underlying algebra is yet to be defined.

Differential interaction nets, introduced by Ehrhard and Regnier in 
\cite{EhrhardRegnier05b}, can be seen as a 
tool for studying non-deterministic 
computation.
Much like linear logic, its syntax comes from semantical observations
and pursues the goal of adding symmetries to logic.
Differential interaction nets can be seen as the programming language
counterpart of
a logic deduced from MELL by adding symmetrized exponential rules :
\emph{co-dereliction}, \emph{co-contraction} and \emph{co-weakening},
and replacing promotion by power series thanks to those rules and
a formal sum structure. This sum is closely related to the additive 
connectives of linear logic and allows to express a non-deterministic
reduction as the sum of the possibilities. Indeed, an encoding of
a finitary fragment of the $\pi$-calculus has been done
by Ehrhard and Laurent in \cite{EhrhardLaurent07a}.

In this paper we construct a GoI for differential 
interaction nets with the aim of
providing an algebraic characterization of paths in non-deterministic
computations. 
\modif{Our focus on differential nets to achieve this goal is motivated
by the existence of a GoI for MELL that we can conservatively extend;
 it is the main reason for not being able to carry directly this work
 in another model of non-determinism.}

In order to deal with the summation of differential interaction
nets we are led to consider a purely syntactic sum. Contrary to
the GoI of MALL we do not recover the original properties of the sum
by mean of variants, but we use our syntactic encoding of the sum to
get information about the execution. We present this GoI by giving an 
equational theory and constructing a concrete model based on reversible
computations. As such we give for the first time an equational theory
suitable for MALL.

This paper is organized in the following way: at first we
will study the abstract notion of paths in interaction nets; 
then we will define differential interaction nets and extend the definition
of paths \modif{to deal with non-determinism}; 
in section 4 we will construct a Geometry of Interaction; and in
a last part we will prove various results of this GoI, from the expected
soundness of the definition to an embedding of the GoI of MELL. 

\section{Paths in interaction nets}

\subsection{Interaction nets}

We set two countable sets,
elements of the first are called \emph{ports}, elements of the second
\emph{symbols}.

An \emph{interaction net} is given by :
    (1) a finite set of \emph{ports};
    (2) a finite set of \emph{cells}: each cell is a finite non-empty
        sequence of pairwise distinct ports and two cells have pairwise
        distinct ports;
    (3) a labelling of cells by symbols;
    (4) a partition of its ports into pairs, called \emph{wires};
    (5) an integer: the number of \emph{loops}.

Ports of a net not present in any cells are called \emph{free ports}.
The first port of a cell is called the \emph{principal port} and the 
$(i+1)$th is called the $i$th \emph{auxiliary port}. For a cell $c$
we will refer to $p(c)$ as the principal port, and $p_i(c)$ as
the $i$th auxiliary port. The number of 
auxiliary ports of a cell is called its \emph{arity}. We suppose
given an arity function $\eta$ from symbols to integers such 
that a cell of symbol $S$ has arity $\eta(S)$.

\begin{figure}
    \centering
    \includegraphics[width=7cm]{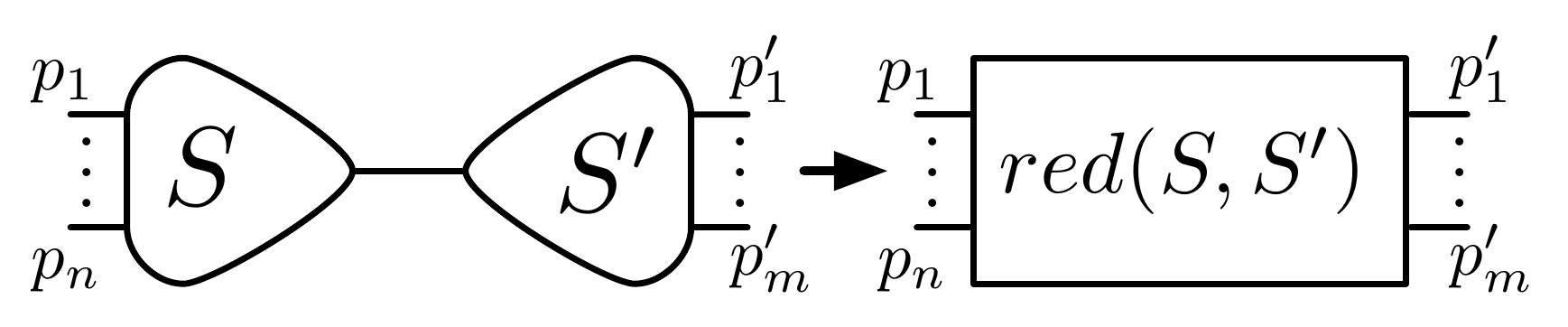}
    \caption{\label{fig:redrule}An interaction rule}
\end{figure}

An \emph{interaction rule} is compound of a pair of symbols $(S,S')$ 
and an interaction net, that we will note $red(S,S')$, 
with a bijection between its free ports and 
$\{1,\cdots,\eta(S)+\eta(S')\}$. A representation of a rule is
given in Fig.~\ref{fig:redrule}, where the bijection is given by the labelling
of auxiliary ports.
Such rule can be made into a reduction
on interaction net by replacing two cells, labelled respectively $S$ and $S'$,
and linked by a wire between their principal ports, with the net 
$red(S,S')$. Proper rewiring is made according to the bijection in the rule;
\modif{
the case of two wires whose ports are pairwise identified corresponds
to an incrementation of the number of loops.
}
We call such reduction an \emph{instance} of the rule.

If we consider a subset of rules such that there is only one rule for
any pair of symbols, then the corresponding reduction is strongly
confluent, also called one-step confluent.

\subsection{Paths}
Let $R$ be an interaction net, we construct a graph $G(R)$ over its ports
with
\begin{itemize}
    \item an undirected edge, called \emph{wire edge},
        between $p$ and $p'$ for any wire $\{p,p'\}$
    \item for every auxiliary port $p_i(c)$, two directed edges,
        called \emph{cell edges}:
        $(p_i(c),p(c))$, noted $c_i$, and $(p(c),p_i(c))$, noted $c_i^r$
\end{itemize}

A path in $R$ is a finite sequence $(p_1, e_1, p_2, e_2,
\cdots, e_n, p_{n+1})$ such 
that $p_i$ is a port of $R$, $e_i$ an edge of $G(R)$ linking $p_i$ and 
$p_{i+1}$ and such that $e_i$ and $e_{i+1}$ are two composable edges
of distinct nature, that is to say there is exactly one wire edge among
them.

Noticing that internal ports in a path are already given by the composability 
condition on edges, and initial and final ports are superfluous when a path
has at least one non-wire edge, we will use the word $e_1 \cdots e_n$
to refer to paths when it is unambiguous.

We note $\paths{R}$ the, possibly infinite, set of paths in $R$,
$\pathsfree{R}$ the subset of paths starting and ending on free ports,
and $\pathsports{R}{p_i}{p_j}$ the set of paths starting at $p_i$ and
ending at $p_j$. For a path $\varphi$ we note $\varphi^r$ its reverse.

\subsection{Path reduction}
\label{sec:pathred}
Let $\mathcal{R}$ be an instance of an interaction rule, 
with the notations of Fig.~\ref{fig:redrule}, and let $R$
be an interaction net where the reduction can be applied.
We call $c$ (resp. $c'$) the cell of symbol $S$ (resp. $S'$).
We can associate to any path in $R$ the set of all paths deduced from it by
replacing every sub-path from $c_i$ to $c'_j$, or $c'_j$ to $c_i$, by
a path in $red(S,S')$ linking these two ports.

To define precisely this association, we have to be sure that 
the path we are considering is not starting or ending with an unfinished
crossing.

\begin{definition}
    Let $R$ be a net, $\mathcal{R}$ a reduction rule applied on
    $c$ and $c'$, a path $\varphi$ is said to be \emph{long enough}
    with respect to $\mathcal{R}$, when $\varphi$ neither starts nor
    ends on the principal ports of $c$ and $c'$.
\end{definition}

The subset of $\paths{R}$ of long enough paths with respect to 
$\mathcal{R}$ is denoted $\pathslong{\mathcal{R}}{R}$.
Remark that a path linking two free ports of a net is long enough 
w.r.t. every reduction, thus $\pathsfree{R} \subseteq
\pathslong{\mathcal{R}}{R}$.

We denote $\crossings{\mathcal{R}}{\varphi}$ the number of 
crossings of $c$ and $c$' 
in $\varphi$, that is the number of its sub-paths of the form
$c_i w {c'_j}^r$ or $c'_j w {c_i}^r$.

\begin{definition}
    \label{def:pathred}
    Let $R$ and $R'$ be two nets such that $R \rightarrow_{\mathcal{R}} R'$,
    we define a mapping $\delta_\mathcal{R}$ from $\pathslong{\mathcal{R}}{R}$
into $\paths{R'}$ with
\begin{eqnarray*}
\modif{\pathred{\mathcal{R}}{\varphi}} &=& 
\modif{\{\varphi\} \textrm{ when } 
\crossings{\mathcal{R}}{\varphi}=0} \\
\pathred{\mathcal{R}}{c_i w {c'_j}^r} &=& 
\pathsports{red(S,S')}{p_i}{p'_j} \\
\pathred{\mathcal{R}}{\varphi^r} &=& 
\{\varphi'^r\ |\ \varphi' \in \pathred{\mathcal{R}}{\varphi}\} \\
\pathred{\mathcal{R}}{\varphi_1 \varphi_2} &=& 
\{ \varphi'_1 \varphi'_2\ |\ 
\varphi'_1 \in \delta_\mathcal{R}(\varphi_1),\ 
\varphi'_2 \in \delta_\mathcal{R}(\varphi_2)\} \\
& & \textrm{when } \crossings{\mathcal{R}}{\varphi_1 \varphi_2} =
\crossings{\mathcal{R}}{\varphi_1} + 
\crossings{\mathcal{R}}{\varphi_2}
\end{eqnarray*}
\end{definition}

Note that in the last case $\delta_\mathcal{R}(\varphi_1 \varphi_2)
\sim \delta_\mathcal{R}(\varphi_1) \times \delta_\mathcal{R}(\varphi_2)$
and the condition on $\crossings{\mathcal{R}}{\varphi_1 \varphi_2}$ means
that the composition occurs outside of the rule pattern.

This function can be seen as some kind of reduction focused on paths, 
and as such
we are interested in the invariant of this pseudo-reduction. 

We set $\delta_{\mathcal{R}_2} \circ \delta_{\mathcal{R}_1}(\varphi)
= \bigcup_{\varphi' \in \delta_{\mathcal{R}_1}(\varphi)} 
\delta_{\mathcal{R}_2}(\varphi')$. 
Thanks to strong confluency we have 
\begin{equation}
    \label{eqn:pathredcomm}
    \delta_{\mathcal{R}_1} \circ \delta_{\mathcal{R}_2} 
    = \delta_{\mathcal{R}_2} \circ \delta_{\mathcal{R}_1}
\end{equation}
when
$\mathcal{R}_1$ and $\mathcal{R}_2$ are both possible reduction of
the same net. With this property saying that a path is invariant by all
reduction is the same as saying that a path is invariant by one
reduction to normal form, and as such the following definition
is in fact independent of the reduction chain.

\begin{definition}
Let $R$ be an interaction net and 
$$R \rightarrow_{\mathcal{R}_1} R_1 \cdots \rightarrow_{\mathcal{R}_n} R_n
$$
be a reduction to normal form, 
$\varphi \in \paths{R}$ is \emph{persistent} if and only if 
$$
\delta_{\mathcal{R}_n} \circ \cdots \circ \delta_{\mathcal{R}_1} (\varphi) \neq 
\emptyset$$
\end{definition}

The function $\delta_R$ satisfies an important property allowing us
to consider it as a path reduction :
\begin{fact}
    \label{fact:pathred}
    Let $R \rightarrow_\mathcal{R} R'$ be a reduction,
    we have
    $$\forall \varphi_1\neq\varphi_2 \in \pathslong{\mathcal{R}}{R},
    \delta_\mathcal{R}(\varphi_1) \cap \delta_\mathcal{R}(\varphi_2)
    = \emptyset$$
    $$\forall \varphi' \in \paths{R'}, \exists \varphi \in \paths{R},
    \varphi' \in \delta_\mathcal{R}(\varphi)$$
\end{fact}

\modif{
For detailed proofs of these properties of $\delta_\mathcal{R}$ one can refer
to \cite{RegnierThesis92} where they are carried in the context of MELL 
proofnets.
}

\section{Differential interaction nets}
\subsection{Definition}
We briefly recall the definition of differential interaction nets, 
for a complete introduction one can refer to \cite{EhrhardRegnier05b}.

\begin{figure*}
    \centering
    \includegraphics[width=13cm]{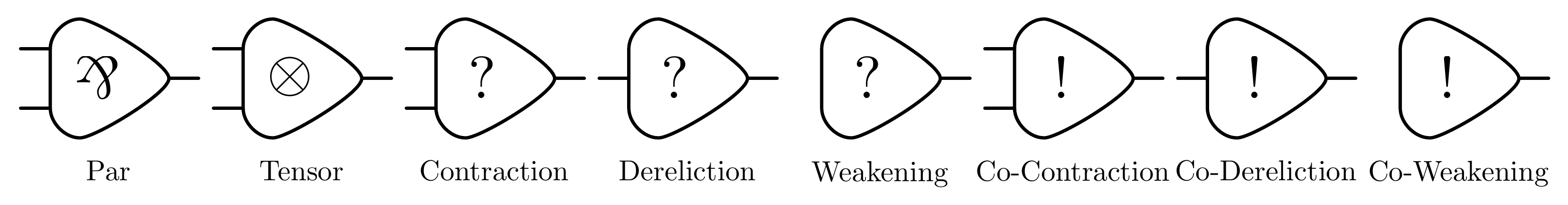}
    \caption{\label{fig:dincells}Differential Interaction Nets cells}
\end{figure*}

\begin{figure*}
    \centering
    \includegraphics[width=13cm]{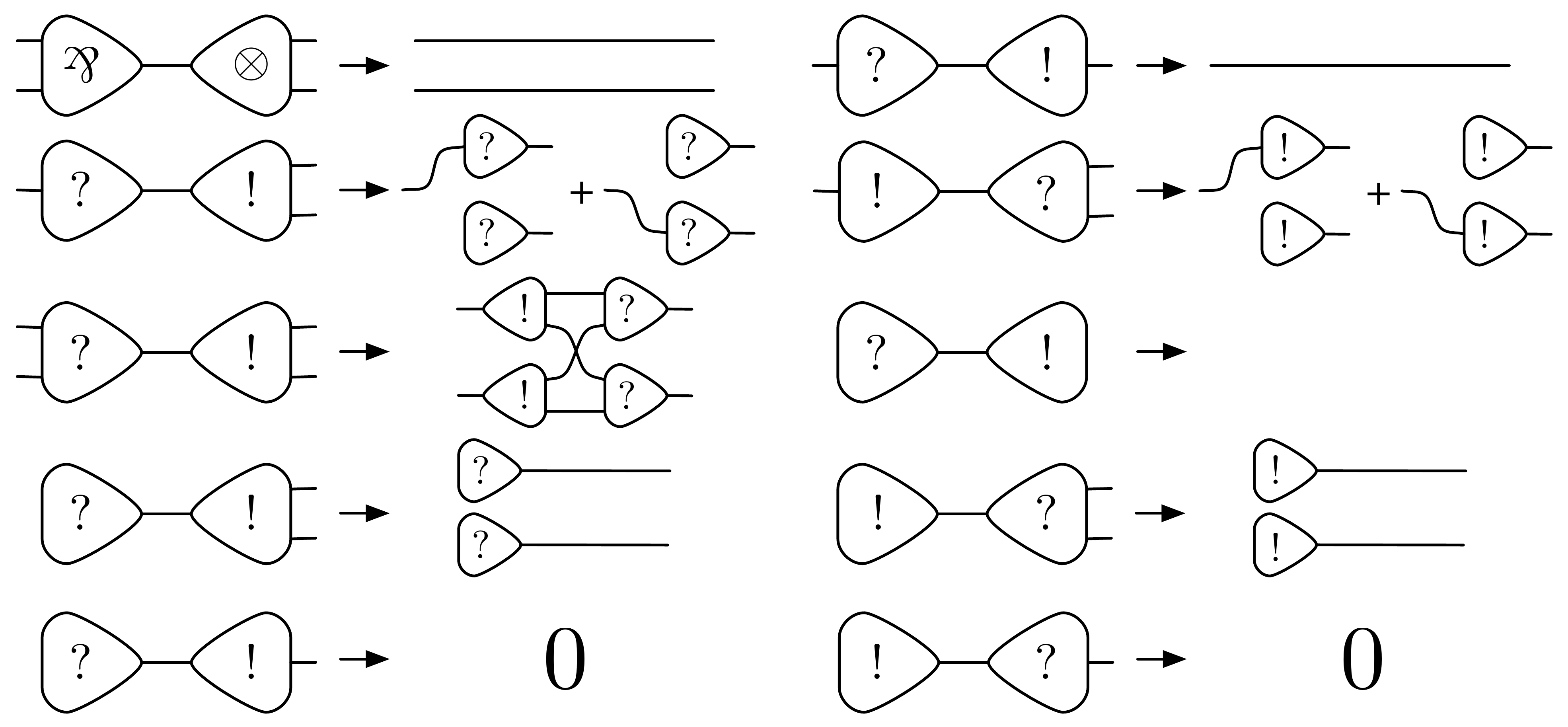}
\caption{\label{fig:dinrules}Differential Interaction Nets reduction rules}
\end{figure*}

Differential interaction nets, \emph{din} for short,
can be seen as a special case of 
interaction nets whose cells are shown in figure~\ref{fig:dincells} and 
reduction rules in figure~\ref{fig:dinrules}, together with a module-like
construction allowing to talk about sums of nets. 



We will call \emph{simple net} a din that is not a sum.

\subsection{Sum-trees}
To distinguish a path in a din $R$ from the same path viewed as
a path in a sum $R+R'$ 
we will define a variant of the notion of din :
we will consider sum of nets as trees with simple nets at their leaves.
Note that this is a syntactic sum, which is non-associative, non-commutative
and does not have a neutral element.

A path in a sum of nets will be of the form $\phi^r \varphi \phi$ where
$\phi$ is a branch of the sum-tree directed from leaf to root, 
$\phi^r$ the reversed branch and $\varphi$ a path in the simple net of the leaf.
We will use the notation $+_1$ (resp. $+_2$) for the left (resp. right) branch
of the sum.

\subsection{Named cells}
Looking at the reduction rules of Fig.~\ref{fig:dinrules}, we can see
that derelictions and co-derelictions cells act as some kind of resources:
they are never duplicated by any rule but the two 
(co-)dereliction/(co-)contraction ones, and in this case the copies are 
separated by a sum. Therefore it is natural to precisely distinguish 
and track them along a reduction.

Let $\names$ be a fixed countable set, we will suppose that every din $R$ is
now labelled in this way :
\begin{itemize}
    \item a label function $l_R$ from dereliction and co-dereliction 
        to $\names$ such that $l_R$ is injective on simple nets
    \item a label for each node of the sum tree, 
        we will use the notation $R +_\alpha R'$ for the sum
        and $+_{\alpha,1}$, $+_{\alpha,2}$ for its branches.
\end{itemize}

We modify the dereliction/co-contraction rule and its dual to
deal with this labelling in this way :

$$\includegraphics[width=7cm]{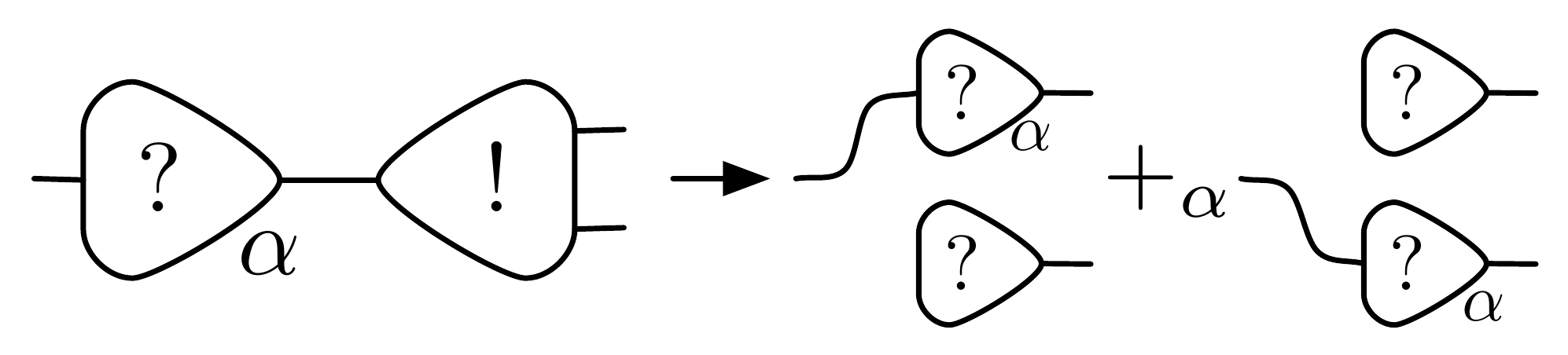}$$

For a din $R$ we denote $\namesr{R}$ the subset of $\names$ of names appearing 
in $R$. Similarly we denote $\namesr{\varphi}$ the names appearing in cells 
traversed by $\varphi$.

\subsection{Relaxed sum-trees}
\modif{To properly extend the definitions of section~\ref{sec:pathred}
we need to}
 relax the structure of sum-trees. 
We consider them quotiented by the relation
\begin{equation}
    \label{eqn:sumselfdistrib}
    \includegraphics[width=6cm]{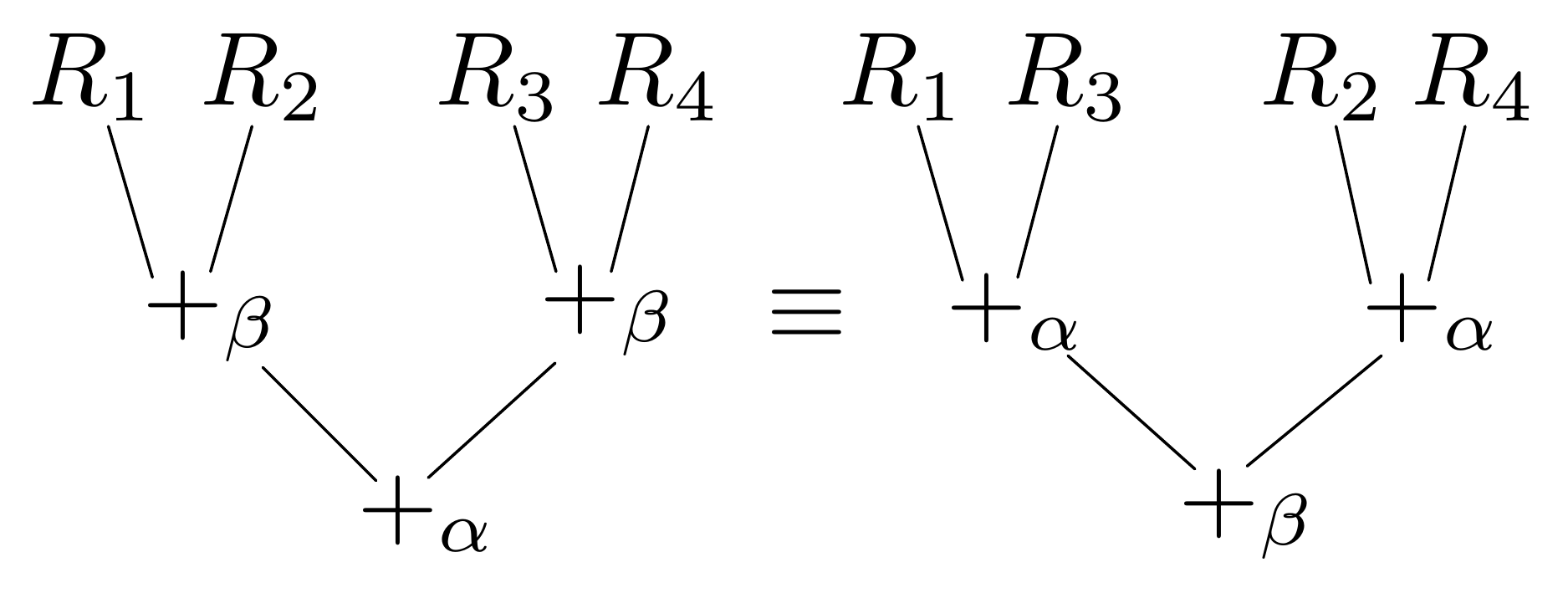}
\end{equation}
where $\alpha \neq \beta$, and the $R_i$s are dins, not necessarily simple.
\modif{
The condition $\alpha \neq \beta$ distinguishes $+_\alpha$ and $+_\beta$
as binary operations, in this context this relation is just an instance of
the \emph{middle four interchange law} (see \cite{MacLane}).
}

This leads us to consider branches of sum-tree also as quotiented by the 
relation 
\begin{equation}
    \label{eqn:branchselfdistrib}
    +_{\alpha,i} +_{\beta,j} \equiv +_{\beta,j} +_{\alpha,i}\quad 
    \forall \alpha \neq \beta \in \names\ \forall i,j \in \{1,2\}
\end{equation}
In the following we will refer to whole classes by a single element when
it is not ambiguous.

This relation is one of the key property of our treatment of the sum. It
means that we have labelled the non-deterministic choices by the precise
context of their appearance, and thanks to this labelling we are able
to say that two choices are independent when their context is distinct.

We will show that the GoI gives a nice algebraic treatment of this labelling,
similar to the handling of Levy's labels by the GoI of the 
$\lambda$-calculus (see \cite{AspertiDanosLaneveRegnier94}).

\subsection{Path reduction of differential interaction nets}
The definition~\ref{def:pathred} is not directly applicable to 
dins because of rules producing sums. We give here a proper extension.

Let $\varphi$ be a path in a simple net $R$ with $R \rightarrow_\mathcal{R}
R_1 +_\alpha R_2$ by one of these rules.

If $\crossings{\mathcal{R}}{\varphi}=0$, $\varphi$ is still present in 
$R_1$ and $R_2$, and we set 
$$\delta_\mathcal{R}(\varphi) = \{ {+_{\alpha,1}}^r \varphi +_{\alpha,1},
+_{\alpha,2}^r \varphi +_{\alpha,2} \}$$

Otherwise, let $c$ be the (co-)dereliction cell and $c'$
the other one, linked by the wire $w$:
\begin{itemize}
    \item either there is $i \in \{1,2\}$ such that 
    $\varphi \cap c' \subseteq \{c'_i, {c'_i}^r\}$, that is $\varphi$ always
    stays on the same side of the \mbox{(co-)contraction}
    cell and we can
    associate to $\varphi$ a path $\varphi_i$ in $R_i$ by erasing the  
    $c'_i$ and ${c'_i}^r$ and set
    $\pathred{\mathcal{R}}{\varphi_i}
    = \{+_{\alpha,i}^r \varphi_i +_{\alpha,i}\}$
    \item or $\varphi$ cannot be mapped to a path in $R_1 +_\alpha R_2$ 
        and we set $\pathred{\mathcal{R}}{\varphi} = \emptyset$.
\end{itemize}
The fact~\ref{fact:pathred} still holds for this generalization of the
$\delta_\mathcal{R}$ function. Note that the 
equation (\ref{eqn:pathredcomm}) needs (\ref{eqn:branchselfdistrib}) 
to be valid in this context.

\subsection{Weak-reduction and weak-persistence}
In the following we will restrict our analysis to the \emph{local} reduction
rules, that is we will not consider the rule dereliction/co-weakening and its
dual. We call this reduction \emph{weak-reduction} and a path persistent 
with respect to this restriction will be called \emph{weakly-persistent}.

\subsection{Purpose of a Geometry of Interaction for dins}
We seek an algebraic structure $\mathcal{A}$ with a distinguished element
$0$ and an embedding 
$w : \paths{R} \mapsto \mathcal{A}$ such that
a path $\varphi$ is weakly-persistent if and only if $w(\varphi) \neq 0$.

\section{Definition of the GoI}
\subsection{The $\dlstar$ algebra}

We recall here the definition of inverse monoids (see \cite{Petrich84}).

\begin{definition}
	A monoid $M$ is an \emph{inverse monoid with zero}, 
	\emph{imz} for short,
if and only $M$ has
a function $x \mapsto \starred{x}$, the \emph{star}, and an element $0$
with
\begin{eqnarray}
\starred{(u v)} & = & \starred{v} \starred{u} \\
\starred{(\starred{u})} & = & u \\
\label{eqn:imzinv}
u \starred{u} u & = & u \\
\label{eqn:imzidem}
u \starred{u} v \starred{v} & = & v \starred{v} u \starred{u} \\
u 0 = &0 u& = 0
\end{eqnarray}
\end{definition}

We will now focus on a special imz that will allow us to weigh paths.

We note $\idem{a} = a \starred{a}$ the idempotent deduced from $a$.
We say that $a$ and $b$ are \emph{orthogonal}, denoted $a \orth b$,
when $\idem{a} \idem{b} = 0$, and that they are
\emph{fully orthogonal}, denoted $a \dorth b$, when they are orthogonal and
$\starred{a} a = \starred{b} b = 1$.
We will note $a \comm b$ to say that $a$ and $b$ commutes;
$a \comms b$ that $a$ commutes with $b$ and $\starred{b}$.
We extend these notations to sets of elements to say that they commute
two by two.

\begin{definition}
    \label{def:dlstar}
Let $\dlstar$ be the inverse monoid freely generated by
$$\{\pop, \qop, \rop{\oc}, \sop{\oc}, \rop{\wn}, \sop{\wn}\}
\cup \bigcup_{\alpha \in \names} \{ \dop{\alpha}{\oc}, \dop{\alpha}{\wn},
\uop{\alpha}, \vop{\alpha}, \eop{\alpha} \}$$
with relations
\begin{eqnarray}
\label{eqn:first_relation}
\label{eqn:pqdorth}
\pop & \dorth & \qop \\
\label{eqn:uvdorth}
\uop{\alpha} & \dorth & \vop{\alpha} \\
\label{eqn:eproj}
\eop{\alpha} \eop{\alpha} &=& \eop{\alpha} \\
\label{eqn:rscomm}
\{ \rop{\oc}, \sop{\oc} \} & \comm & \{ \rsop{\wn}, \ssop{\wn} \} 
\end{eqnarray}
\begin{equation}
\label{eqn:uvcomms}
\begin{array}{l}
\{ \uop{\alpha}, \vop{\alpha}, \eop{\alpha} \}  \comms
\{ \pop, \qop, \rop{\wn}, \sop{\wn}, \rop{\oc}, \sop{\oc} \} \cup\\ 
\quad\quad\quad \bigcup_{\beta \in \names - \{ \alpha \}} 
 \{ \dop{\beta}{\wn}, \dop{\beta}{\oc},
\uop{\beta}, \vop{\beta}, \eop{\beta} \}
\end{array}
\end{equation}
\begin{eqnarray}
\label{eqn:drred}
\dsop{\alpha}{t} \rop{t'} & = & \uop{\alpha} \dsop{\alpha}{t} \usop{\alpha} \\
\label{eqn:dsred}
\dsop{\alpha}{t} \sop{t'} & = & \vop{\alpha} \dsop{\alpha}{t} \vsop{\alpha} \\
\label{eqn:dinv}
\label{eqn:last_relation}
\dsop{\alpha}{\oc} \dop{\beta}{\wn} & = & \eop{\alpha} \eop{\beta} 
\end{eqnarray}
where $\alpha, \beta \in \names$, $\alpha \neq \beta$ and $t, t' \in \{\wn,\oc\}$,
$t \neq t'$.
\end{definition}


%

\modif{
The equation (\ref{eqn:dinv}) could not be replaced by the simpler
$\dsop{\wn}{\alpha} \dop{\oc}{\beta} = 1$ as all our models satisfies
unicity of left inverse and it leads to a collapse of generators, leading
eventually to a persistent path of weight $0$.
}

The last three relations are summed up in the following diagram,
\modif{where $\leftrightarrow$ stands for \emph{dual} and 
$\rightarrow$ stands for \emph{reduces to}}.
$$
\begin{diagram}[size=1em]
    \rop{\wn} && \rCorresponds && \rop{\oc}\\
    &\rdTo & & \ldTo&\\
    && \uop{\alpha} &&
\end{diagram}
\quad
\begin{diagram}[size=1em]
    \sop{\wn} && \rCorresponds && \sop{\oc}\\
    &\rdTo & & \ldTo&\\
    && \vop{\alpha} &&
\end{diagram}
\quad
\begin{diagram}[size=1em]
    \dop{\alpha}{\wn} && \rCorresponds && \dop{\alpha}{\oc}\\
    &\rdTo & & \ldTo&\\
    && \eop{\alpha} &&
\end{diagram}
$$
Up to these relations, we have five generators : 
the multiplicatives $p$ and $q$, and these three.

We distinguish two sub-monoids of $\dlstar$ :
\begin{itemize}
    \item $\dlstarpme$ the monoid generated by $\pop,\qop,
        \rop{\wn},\sop{\wn},\rop{\oc},\sop{\oc}, 
        \dop{\alpha}{\wn}, \dop{\alpha}{\oc},\ldots$
    \item $\dlstarpa$ the monoid generated by $\uop{\alpha},
        \vop{\alpha}, \eop{\alpha}, \ldots$
\end{itemize}

We ensure that definition~\ref{def:dlstar} is non-trivial by constructing
a proper model, that is an inverse monoid
satisfying  equations (\ref{eqn:first_relation}--\ref{eqn:last_relation}) 
where $0 \neq 1$. Such construction is made in~\ref{sec:model}.

\subsection{Weighting paths}

\begin{figure}
    \centering
    \includegraphics[width=6cm]{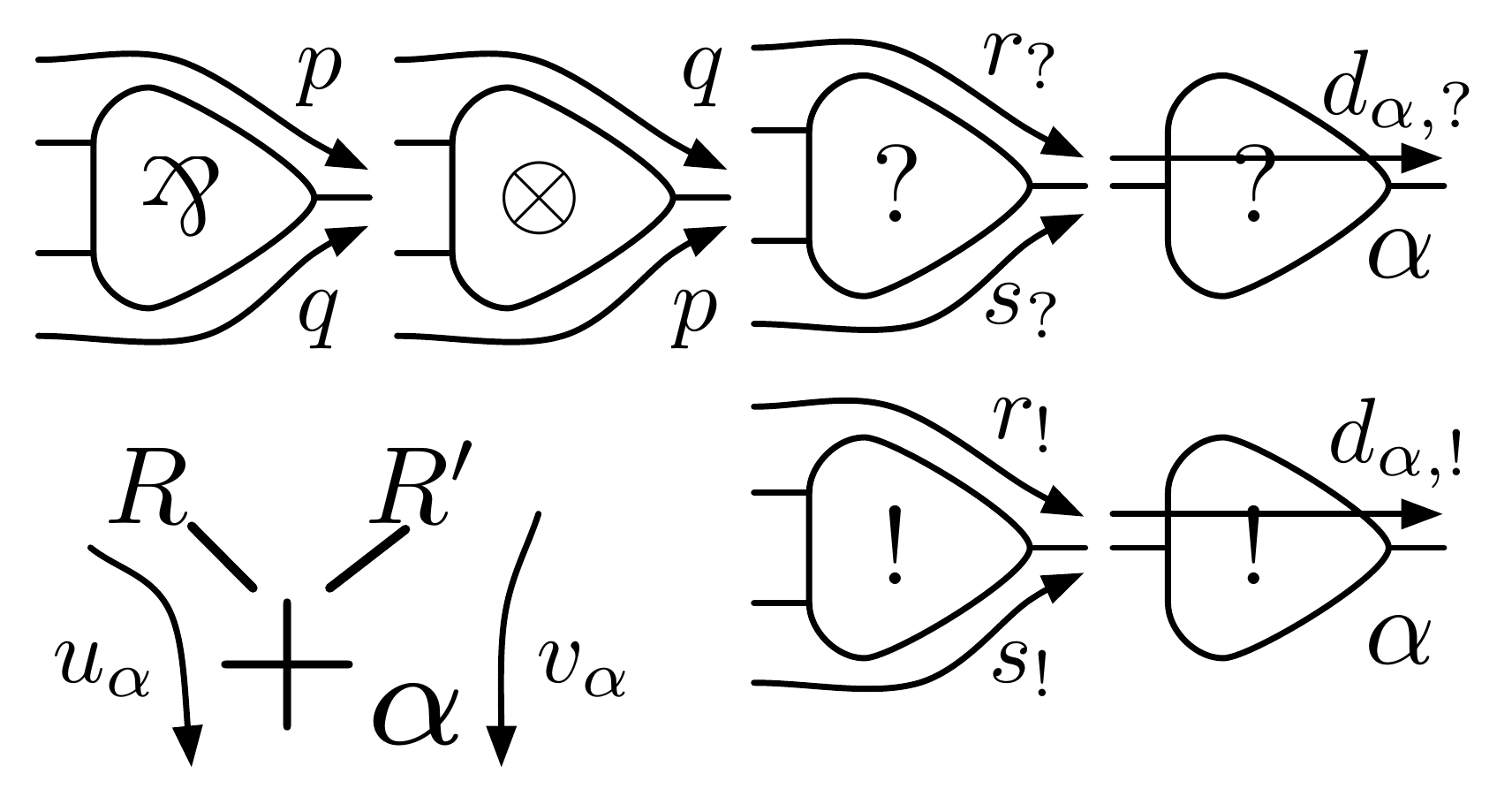}
    \caption{\label{fig:pathweight}Weighting of cell-edges}
\end{figure}

Let $R$ be a din and $\varphi$ a path in $R$.
If $\varphi = e_1 \cdots e_n$ we define $w(\varphi) = w(e_n) \cdots
w(e_1)$ with $w(e)=1$ for a wire $e$, $w(c_i)$ and 
$w(+_{\alpha,i})$ given by Fig.~\ref{fig:pathweight} 
and $w(e^r) = \starred{w(e)}$. 

\modif{
We have to ensure that this definition is independent from the
choice of the representation of relaxed sum-trees.
\begin{fact}
    Let $R$ and $R'$ be two dins with $R \equiv R'$,
    $\varphi \in \mathcal{P}(R)$ and $\varphi' \in \mathcal{P}(R')$
    with $\varphi \equiv \varphi'$, we have
    $$w(\varphi) = w(\varphi')$$
\end{fact}
}

\modif{
\begin{proof}
    Direct by observing that (\ref{eqn:uvcomms}) is the algebraic
    counterpart of (\ref{eqn:branchselfdistrib}).
\end{proof}
}


We will call \emph{regular} a path
$\varphi$ such that $w(\varphi) \neq 0$. 

When reasoning along a reduction we want to keep track of deleted names,
occurring in a dereliction/co-dereliction rule, and algebraically this is
the role of $e_\alpha$. Let $\varphi$ be a path in a din $R$ and 
$\varphi'$ a path in a din $R'$, we define $n_{\varphi'}(\varphi)$, the
\emph{normalizing factor} of $\varphi$ with respect to $\varphi'$, to be
$$n_{\varphi'}(\varphi) = \prod_{\alpha \in \namesr{\varphi'} - 
\namesr{\varphi}}
\eop{\alpha}$$
Similarly, we define
$$n_R(\varphi) = \prod_{\alpha \in \namesr{R} - \namesr{\varphi}} 
\eop{\alpha}$$

We have the following lemma relating normalizing factors along a reduction :
\begin{lemma}
    \label{lemma:normred}
    Let $R$ be din, $R \rightarrow_\mathcal{R} R'$ 
    be a reduction, $\varphi$ be a path
    in $\paths{R}$ and suppose that 
    $\delta_{\mathcal{R}}(\varphi)= \{\varphi'\}$, then 
    $$n_R(\varphi') = n_R(\varphi) n_{\varphi}(\varphi')$$
\end{lemma}

\begin{proof}
    The proof is trivially induced by the fact that 
    $$\namesr{\varphi'} \subseteq \namesr{\varphi}$$
\end{proof}

\subsection{A $\dlstar$ realization : $\Mdlstar$ }
\label{sec:model}
Let $\Sigma^\star$ be the set of words over $\Sigma$, we view it
as the set of stacks containing elements of $\Sigma$, where the empty stack
corresponds to the empty word, and concatenation of stacks to concatenation
of words. 

We call \emph{token} an element of 
$$\tokens = \words{\bool} \times \words{(\words{\bool})} 
\times \words{(\words{\bool})} \times \finfuncset{\names}{\words{\bool}}$$
where the fourth set is the subset of $(\words{\bool})^\names$ of functions
almost everywhere equal to the empty word.

For a given token $t = (M, E_\wn, E_\oc, f) \in \tokens$
 we will refer to the stack $M$ as
the \emph{multiplicative stack}, to the stack $E_\wn$ (resp. $E_\oc$) as the
\emph{$\wn$-exponential stack} (resp. \emph{$\oc$-exponential stack}) and 
to the function $f$ as the \emph{index function}. We use the notation
$\funextends{\idxf}{\alpha}{\sigma}$ for the function such that 
$$
\funextends{\idxf}{\alpha}{\sigma}(\beta) = 
\left\lbrace
\begin{array}{ll}
    \sigma & \textrm{ if } \beta = \alpha \\
    \idxf(\beta) & \textrm{ otherwise }
\end{array}
\right.
$$
We call operation a partial invertible function from tokens to tokens.
We note $1$ the identity and $0$ the function of empty domain.
We define the following operations :
\begin{eqnarray*}
\dtoken &\operation{\Mpop}& \token{\bzero \concat \ms}{\ewns}{\eocs}{\idxf} \\
\dtoken &\operation{\Mqop}& \token{\bone \concat \ms}{\ewns}{\eocs}{\idxf} \\
\token{\ms}{\sigma \concat \ewns}{\eocs}{\idxf} 
&\operation{\Mrop{\wn}}& 
\token{\ms}{(\bzero \concat \sigma) \concat \ewns}{\eocs}{\idxf} \\
\token{\ms}{\sigma \concat \ewns}{\eocs}{\idxf} 
&\operation{\Msop{\wn}}& 
\token{\ms}{(\bone \concat \sigma) \concat \ewns}{\eocs}{\idxf} \\
\dtoken &\operation{\Mdop{\alpha}{\wn}}& 
\token{\ms}{\epsilon \concat \ewns}{\sigma \concat \eocs}{\idxf}  
\\ & & \quad \textrm{where } \idxf(\alpha) = \sigma \\
\dtoken &\operation{\Muop{\alpha}}& 
\token{\ms}{\ewns}{\eocs}{\funextends{\idxf}{\alpha}{\bzero \concat \sigma}}
\\ & & \quad \textrm{where } \idxf(\alpha) = \sigma \\
\dtoken &\operation{\Mvop{\alpha}}& 
\begin{array}{l}
\token{\ms}{\ewns}{\eocs}{\funextends{\idxf}{\alpha}{\bone \concat \sigma}}
\\
\quad \textrm{where } \idxf(\alpha) = \sigma
\end{array}
\\
\dtoken \textrm{ with} &\operation{\Meop{\alpha}}& \dtoken \\ 
\idxf(\alpha) = \epsilon &&
\end{eqnarray*}
For every operation $\mathfrak{g} = (\mathfrak{g}_m, \mathfrak{g}_\wn,
\mathfrak{g}_\oc, \mathfrak{g}_f)$ we can define its dual
$\Delta(\mathfrak{g}) =  (\mathfrak{g}_m, \mathfrak{g}_\oc,
\mathfrak{g}_\wn, \mathfrak{g}_f)$.
We set $$\Mrop{\oc}=\Delta(\Mrop{\wn}), \Msop{\oc}=\Delta(\Msop{\wn})
\textrm{ and } \Mdop{\alpha,\oc}=\Delta(\Mdop{\alpha,\wn})$$
Let $\Mdlstar$ be the smallest set containing these operations and
closed under composition and inversion. It has a natural structure
of imz with composition, inversion as the star,
identity as $1$, and the nowhere defined operation as $0$.

\begin{fact}
	$\Mdlstar$ is an imz satisfying 
	equations (\ref{eqn:first_relation}--\ref{eqn:last_relation}). 
\end{fact}

\begin{proof}
For the equations (\ref{eqn:pqdorth}) and 
(\ref{eqn:uvdorth}), orthogonality is implied by the fact 
that $\bzero \neq \bone$ and full orthogonality by the fact
that $\Mpop, \Mqop, \Muop{\alpha}$ and $\Mvop{\alpha}$ have full
domain. The equation (\ref{eqn:eproj}) is trivial.

The commutation relations (\ref{eqn:rscomm}) are deduced from the
fact that $\Mrop{t}$ and $\Msop{t}$ only act on the stack $\es{t}$
; the relations (\ref{eqn:uvcomms}) from the focusing of $\Muop{\alpha}$
and $\Mvop{\alpha}$ on the value of $\idxf$ in $\alpha$.

We only need to check \ref{eqn:drred} and we will get \ref{eqn:dsred}   
by symmetry. We will prove an equivalent equation :
$$\Mrsop{\wn} \Mdop{\alpha}{\oc} = \Muop{\alpha} \Mdop{\alpha}{\oc}
\Musop{\alpha}$$
The operation $\Mrsop{\wn} \Mdop{\alpha}{\oc}$ and $\Muop{\alpha}
\Mdop{\alpha}{\oc} \Musop{\alpha}$ are only defined on tokens 
where $\idxf(\alpha) = \bzero \concat \sigma$ and for those : 
$$
\begin{array}{rl}
&\Mrsop{\wn} \Mdop{\alpha}{\oc} (\ms, \ewns, \eocs, \idxf)\\ 
=&\Mrsop{\wn} (\ms, (\bzero \concat \sigma) \concat \ewns,
\epsilon \concat \eocs, \idxf)\\
=&(\ms, \sigma \concat \ewns, \epsilon \concat \eocs, \idxf) 
\end{array}
$$

$$
\begin{array}{rl}
&\Muop{\alpha} \Mdop{\alpha}{\oc} \Musop{\alpha} \dtoken\\
=&\Muop{\alpha} \Mdop{\alpha}{\oc}
(\ms, \ewns, \eocs, \funextends{\idxf}{\alpha}{\sigma})\\ 
=&\Muop{\alpha} (\ms, \sigma \concat \ewns, \epsilon \concat
\eocs, \funextends{\idxf}{\alpha}{\sigma})\\ 
=&(\ms, \sigma \concat \ewns, \epsilon \concat \eocs, \idxf) 
\end{array}
$$
The last equation to check is (\ref{eqn:dinv}). We have
$$
\begin{array}{rl}
&\Mdsop{\alpha}{\oc} \Mdop{\beta}{\wn} \dtoken\\
=&\Mdsop{\alpha}{\oc} (\ms, \epsilon \concat \ewns, \idxf(\beta) \concat
\eocs, \idxf) \\
=& \dtoken
\end{array}
$$
for tokens where $\idxf(\alpha) = \idxf(\beta) = \epsilon$; 
while the operation $\Meop{\alpha} \Meop{\beta}$ is the identity
restricted to tokens satisfying this condition.
\end{proof}

In $\Mdlstar$ we have $0 \neq 1$ and $\Meop{\alpha} \neq \Meop{\beta}$ 
for $\alpha \neq\beta$, which means that those elements are different
in $\dlstar$ seen as the most general imz satisfying relations 
(\ref{eqn:first_relation}--\ref{eqn:last_relation}).

\section{Results}
\subsection{Soundness}
We prove here that the GoI we have defined suit our purpose,
that is it allows an algebraic characterization of persistent paths.
Almost every proof in this paper will rely on the following 
lemma.

\begin{lemma}[Fundamental lemma]
    Let $R$ be a simple net, 
    $R \rightarrow_\mathcal{R} R'$ a step of weak-reduction.
    For all path $\varphi$ in $R$ 
    long enough with respect to $\mathcal{R}$,
    such that $\crossings{\mathcal{R}}{\varphi} \neq 0$,
    either, 
    $\delta_\mathcal{R}(\varphi) = \emptyset$
     and  $w(\varphi)=0$, or,
    $\delta_\mathcal{R}(\varphi)=\{\varphi'\}$ and  
    $$w(\varphi) = n_\varphi(\varphi') w(\varphi')
    $$
    \label{lemma:onestepsoundness}
\end{lemma}

\begin{figure}
    \centering
    \subfigure[Persistent]{
    \includegraphics[width=8cm]{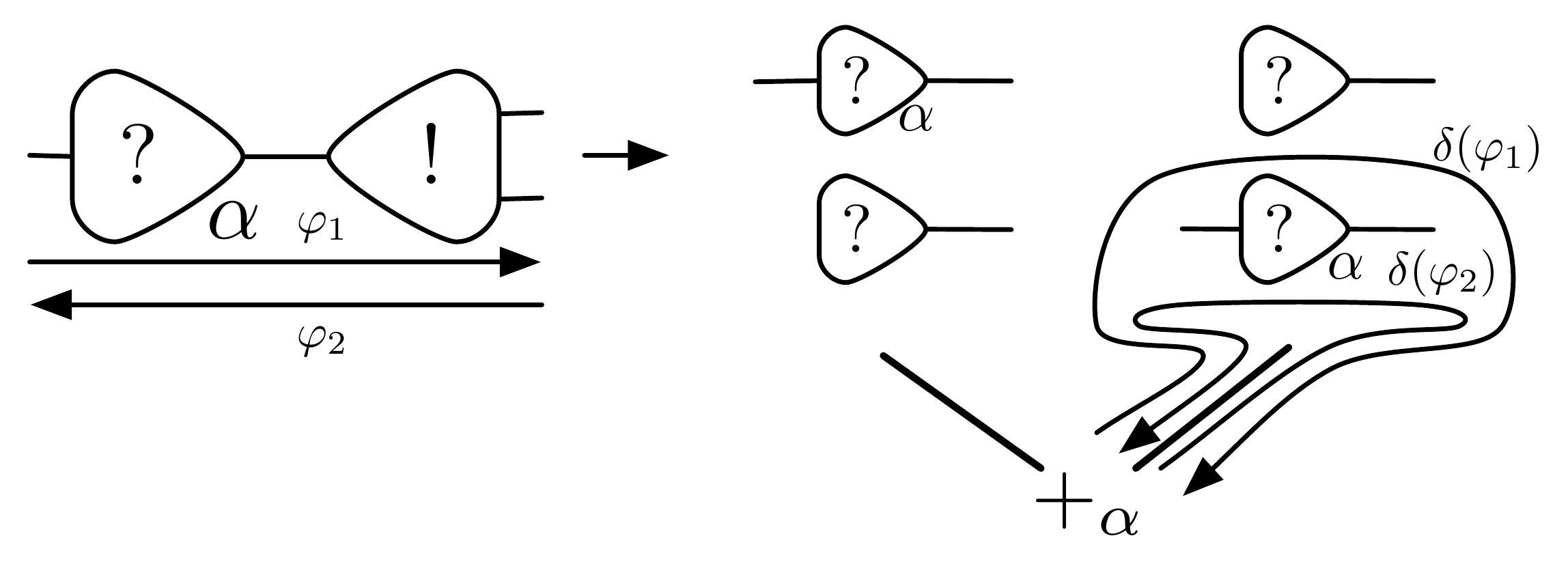}
    }
    \subfigure[Not persistent]{
    \includegraphics[width=8cm]{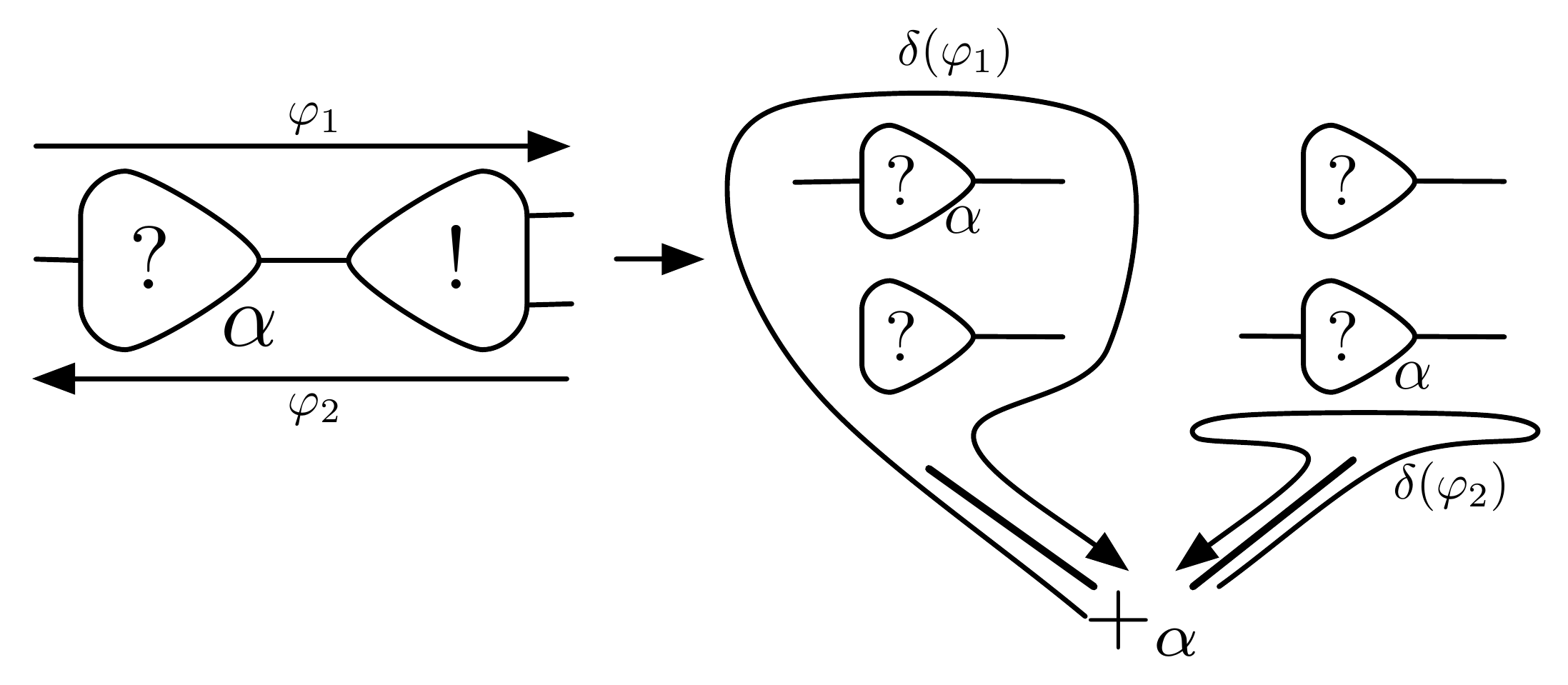}
    }
    \caption{\label{fig:crossings}
    Possibilities for two successive crossings of 
    a reduction \mbox{dereliction/co-contraction}}
\end{figure}

\begin{proof} 
    Before starting the proof, let us remark that the normalizing factors
    are always equal to $1$, except for a rule dereliction/co-dereliction, and 
    in
    this case we have $n_\varphi(\varphi')= e_\alpha e_\beta$ were $\alpha$ is
    the name of dereliction and $\beta$ of the co-dereliction.
 
    We will only do the most important cases here, the other ones being 
    either trivially deduced from the axioms of $\dlstar$ or symmetric to
    these ones.
    
    Let $c$ and $c'$ be the two cells reduced by $\mathcal{R}$ and $e$ 
    the wire linking their principal port, and let
    $\varphi = \varphi_1 \cdots \varphi_n$ be a decomposition
    of $\varphi$ such that, for all $i$, 
    $\crossings{\mathcal{R}}{\varphi_i} = 1$.
    We can further decompose the $\varphi_i$'s :
    $$\varphi_i = \varphi'_i (c_1 e {c'_{m_i}}^r)^{t_i} \varphi''_i$$
    with $m_i \in \{1,2\}$ and $t_i$ is nothing or $r$. 
    
    Suppose as in Fig.~\ref{fig:crossings}, 
    that $c$ is a dereliction of name $\alpha$, $c'$ 
    a co-contraction, and 
    $R' = R_1 +_\alpha R_2$. In this case, thanks to (\ref{eqn:drred}) and
    (\ref{eqn:dsred})
    $$w(\varphi_i) = w(\varphi'_i) a_{m_i}
    (\dop{\alpha}{\wn})^{t'_i} \starred{a_{m_i}} w(\varphi''_i)$$ 
    with $a_1 = \uop{\alpha}$, $a_2 = \vop{\alpha}$ and $t'_i = \star$
    if $t_i = r$. As no weight indexed by $\alpha$ can appear in 
    $w(\varphi'_i)$ or $w(\varphi''_i)$ we have, thanks to
    (\ref{eqn:uvcomms})
    $$w(\varphi_i) = a_{m_i} w(\varphi'_i) 
    (\dop{\alpha}{\wn})^{t'_i} w(\varphi''_i) \starred{a_{m_i}}$$ 
   
    There two possibilities for these successive crossings,
    as it is depicted in Fig.~\ref{fig:crossings}.
    If there is $i_0$ such
    that $m_{i_0} \neq m_{i_0+1}$ , in this case we have, by definition,
    $\pathred{\mathcal{R}}{\varphi} = \emptyset$ and
    $$w(\varphi_{i_0}) w(\varphi_{i_0+1}) = a_{m_{i_0}}
    \omega \starred{a_{m_{i_0}}} a_{m_{i_0+1}} \omega' 
    \starred{a_{m_{i_0+1}}} = 0$$
    by (\ref{eqn:uvdorth}). Otherwise all $m_i$s are equal,
    $\pathred{\mathcal{R}}{\varphi} = \{ +_{\alpha,m_1} \varphi' 
    +_{\alpha,m_1}^r \}$, with $\varphi'$ deduced from $\varphi$ by deleting all
    edges in $c'$, and we have 
    \begin{eqnarray*}
    w(\varphi) &=& a_{m_1} (\prod_i w(\varphi'_i) 
    (\dop{\alpha}{\wn})^{t'_i}
    w(\varphi''_i) ) \starred{a_{m_1}}\\
    &=& a_{m_1} w(\varphi') \starred{a_{m_1}}
    \end{eqnarray*}

    Now, suppose that $c$ is a dereliction of name $\alpha$ and $c'$ 
    a co-dereliction of name $\beta$. Then $\pathred{\mathcal{R}}{\varphi}
    = \{ \varphi'_1 \varphi''_1 \cdots \varphi'_n \cdots \varphi''_n \}
    = \{ \varphi' \}$
    and, by (\ref{eqn:uvcomms}) 
    \begin{eqnarray*}
        w(\varphi) &=& \prod_i w(\varphi'_i) \eop{\alpha} \eop{\beta}
        w(\varphi''_i) \\
        &=& \eop{\alpha} \eop{\beta} \prod_i w(\varphi'_i) w(\varphi''_i)\\
        &=& n_\varphi(\varphi') w(\varphi')
    \end{eqnarray*}
\end{proof}

\begin{theorem}[soundness of the GoI]
    If $R$ is a normalizing din
    then, for all path $\varphi$ in $R$ linking two free ports we have
    $$\varphi \textrm{ weakly-persistent } \iff \varphi \textrm{ regular }$$
    \label{theorem:soundness}
\end{theorem}

\begin{proof}
    By induction on the length of the reduction applying either 
    lemma~\ref{lemma:onestepsoundness} when $\crossings{\mathcal{R}}{\varphi}
    \neq 0$ or using a direct argument.
\end{proof}

\subsection{Structure of the weights of paths}
\begin{figure}
    \centering
    \includegraphics[width=5cm]{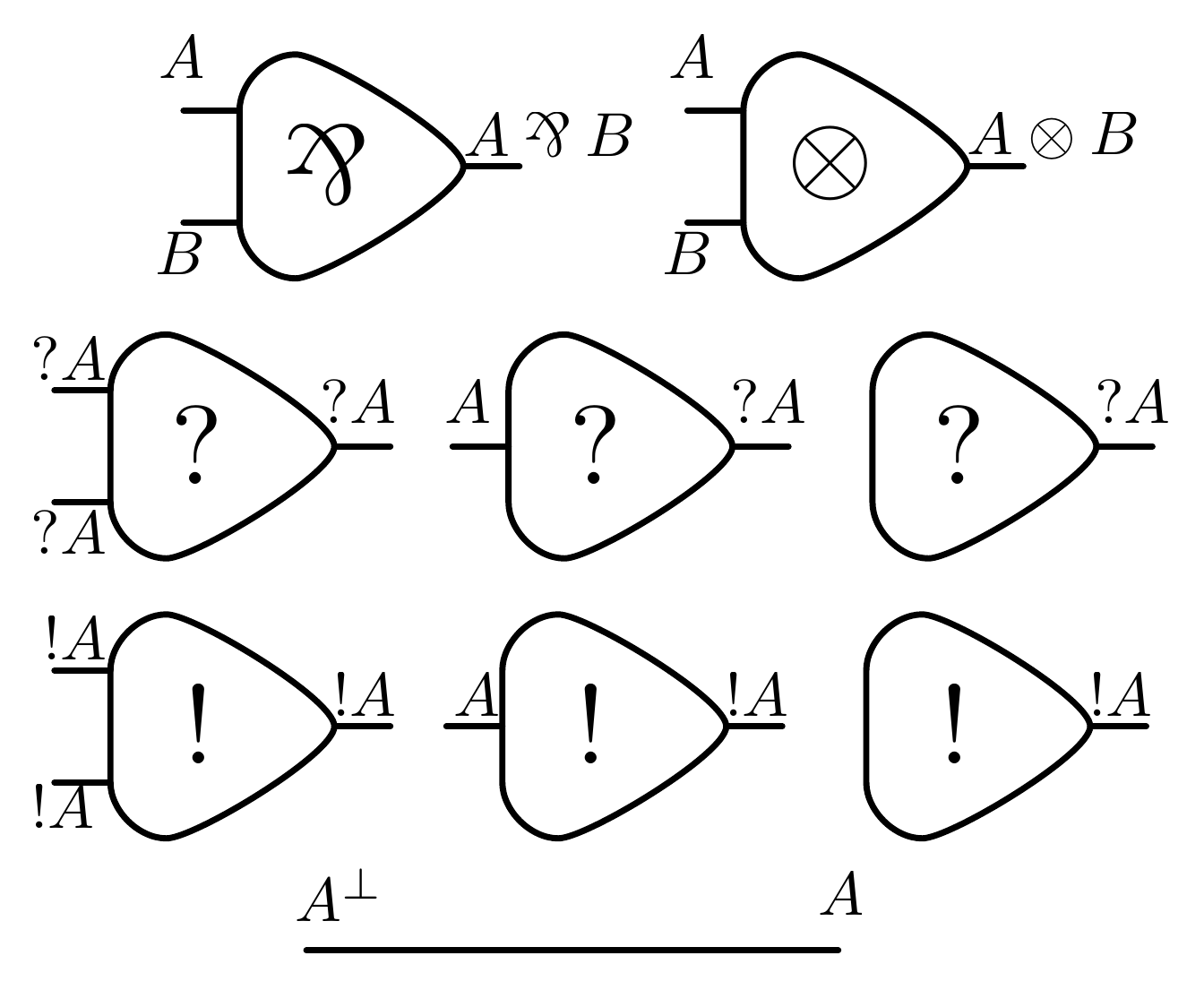}
    \caption{\label{fig:dintype}Typing rules for dins}
\end{figure}

We give here a theorem, adapting the $a \starred{b}$ theorem of
\cite{DanosRegnier95} and \cite{AspertiDanosLaneveRegnier94},
asserting the existence of 
a canonical representation for the weight of a path.
This theorem is only valid for the particular case of typed dins which is 
only interesting: typed means that the program will have a well behaved
reduction.
Typed dins are dins 
with ports labelled by formulae satisfying conditions summed up
in Fig.~\ref{fig:dintype} together with the relations : 
$A \parr B = (A^\bot \otimes B^\bot)^\bot$ and
$\wn A = (\oc A^\bot)^\bot$.
For an in-depth study of this type system see \cite{EhrhardRegnier05b}.

\begin{lemma}
    Let $R$ be a well-typed din and assume we have a reduction
    $R \rightarrow_* R_0$ where $R_0$ is in normal form,
    then each leaf in the sum-tree $R_0$ have no
    wire linking two principal ports.
    \label{lemma:welltypedcutfree}
\end{lemma}

\begin{proof}
    The proof is just an application of subject-reduction and noticing
    that a non reducible wire linking two principal ports would be
    ill-typed.
\end{proof}

\begin{figure}
    \centering
    \includegraphics[width=6cm]{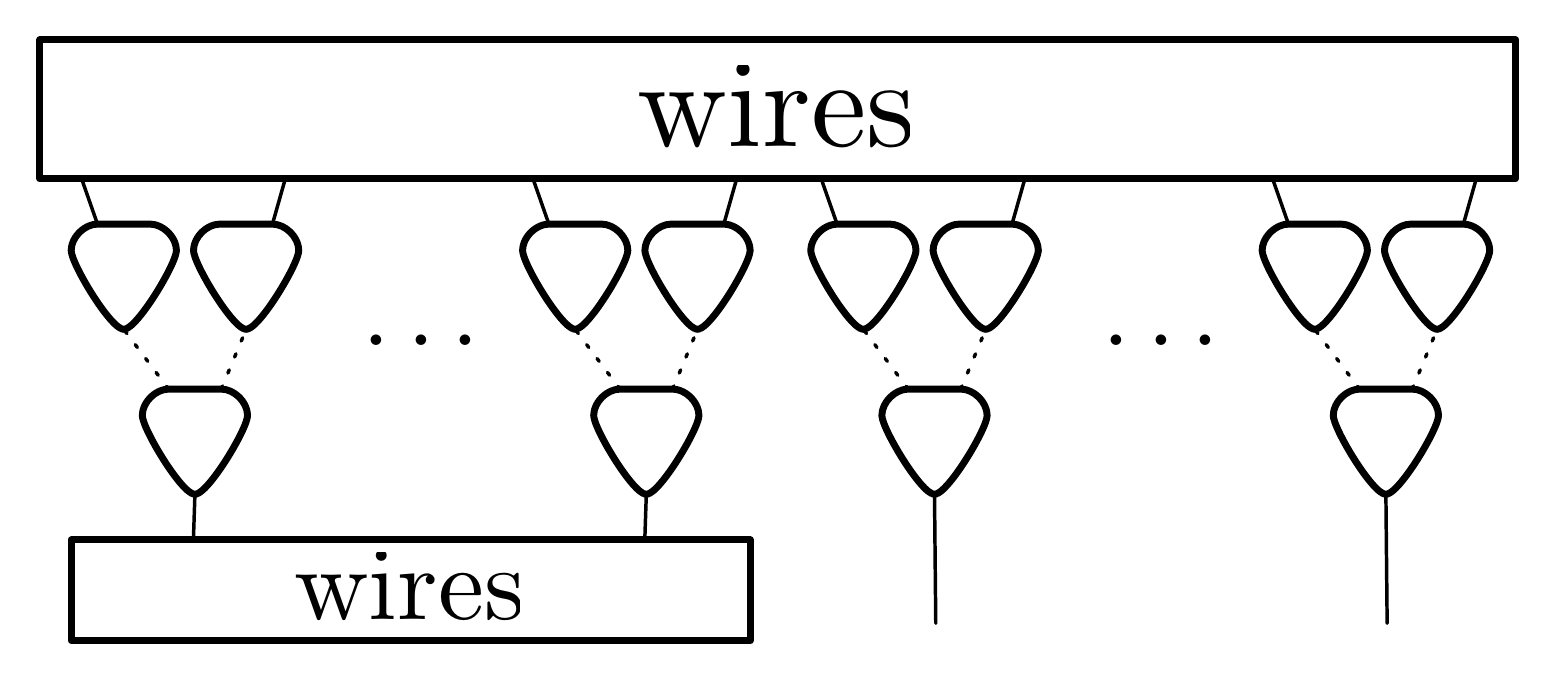}
    \caption{\label{fig:generalnet}The general form of a simple net}
\end{figure}

Looking at the general form of a simple net as it is depicted in
Fig.~\ref{fig:generalnet} and applying the previous lemma, 
we can directly express the weight of paths in a normal din.

The only differences between a normal din and a 
weakly-normal din are the remaining wires linking (co)-weakening
to (co-)dereliction. Wires and (co-)weakening cells being invisible to
weighting we can use the previous lemma to compute the weight of paths in
weakly-normal din in a canonical form.

Thanks to this remark and lemma~\ref{lemma:onestepsoundness},
we can use this form in normalisable net by going back along a reduction.
Thus, we prove the following theorem, which is a generalization
of the usual $a \starred{b}$ form of MELL.

\begin{theorem}[$\alpha a \starred{b} \starred{\alpha}$ form]
    Let $\varphi$ be a path in a well-typed and normalisable din.
    Then either $w(\varphi) = 0$ or 
    $\exists \alpha \in \dlstarpa$,
    $\exists a, b \in \dlstarpme$ such that 
        $$w(\varphi) = \alpha a \starred{b} \starred{\alpha}$$
    \label{theorem:weightform}
\end{theorem}

By summing up these results we have the following fact.
\begin{fact}
    \label{fact:finiteregular}
    Let $R$ be a well-typed and normalisable din, there are
    only finitely many regular paths in $R$.
\end{fact}

The converse of this fact is yet to be proved, but observing that it is
provable in the case of MELL we can probably adapt the proof to this case.
Though, the direct argument on the strictly decreasing length of a path
has to be refined to cope with contraction/co-contraction and 
\mbox{(co-)dereliction/(co-)contraction}.

\subsection{Normalized execution}

When reasoning on paths we can usually get back a global invariant of
reduction by computing the so-called \emph{execution}, defined
for example in \cite{DanosRegnier95} as
$$\EX{R} = \sum_{\varphi \in \pathsfree{R}} w(\varphi)$$
where the summation occurs in a module over weights whose neutral
element for addition is $0$, and we restrict ourselves to cases
where the sum is finite by using the fact~\ref{fact:finiteregular}.

Here we cannot directly reuse this definition because
lemma \ref{lemma:onestepsoundness} only asserts equality up to normalization
factors. 

We define the \emph{normalized execution} with respect 
to a din $R$ of a din $R'$ as
$$\NEX{R}{R'} = \sum_{\varphi \in \pathsfree{R'}} 
n_R(\varphi) w(\varphi)$$

\begin{fact}
    \label{lemma:onestepnex}
    Let $R$ be a well-typed and normalisable
    simple net and $R \rightarrow_\mathcal{R} R'$,
    $$\NEX{R}{R} = \NEX{R}{R'}$$
\end{fact}

\begin{proof}
    The proof is relying on lemma \ref{lemma:onestepsoundness}.
    One case is out of the scope of this lemma and need to be independently
    checked: if the reduction is the application of a
    \mbox{(co-)dereliction/(co-)contraction}
    rule when we have a path $\varphi$ not crossing the pattern, but, if
    $\alpha$ is the (co-)dereliction name, observing that
\begin{equation}
    \label{eqn:uvsum}
    \uop{\alpha} \usop{\alpha} + \vop{\alpha} \vsop{\alpha} = 1
\end{equation}
    and no $\alpha$ indexed weight can appear in $w(\varphi)$, we
    have 
    \begin{equation}
        \label{eqn:laststepsoundness}
    \uop{\alpha} w(\varphi) \usop{\alpha} + \vop{\alpha} w(\varphi)
  \vsop{\alpha} \modif{ = w(\varphi) (\uop{\alpha} \usop{\alpha} +
  \vop{\alpha} \vsop{\alpha})} = w(\varphi)
    \end{equation}
    Now to prove the main result we start by grouping in $\NEX{R}{R'}$
    the weights of paths coming from the same path in $R$ by 
    $\delta_\mathcal{R}$.
    $$\NEX{R}{R'} = 
    \sum_{\varphi \in \pathsfree{R}} \sum_{\varphi' \in 
    \delta_\mathcal{R}(\varphi')} n_R(\varphi') w(\varphi')$$
    
    If $\crossings{\mathcal{R}}{\varphi} \neq 0$, we have by 
    lemma \ref{lemma:onestepsoundness}, either $\delta_R(\varphi) =
    \emptyset$, and $\varphi$ has no contribution to this sum, or
    $\delta_R(\varphi) = \{\varphi'\}$ with
    \begin{eqnarray*}
    n_R(\varphi') w(\varphi') &=& n_R(\varphi) n_\varphi(\varphi')
    w(\varphi') \textrm{ by lemma \ref{lemma:normred}}\\
    &=& n_R(\varphi) w(\varphi) \textrm{ by lemma \ref{lemma:onestepsoundness}}
    \end{eqnarray*}
    Otherwise, either the rule is not a 
    \mbox{(co-)dereliction/(co-)contraction}
    and $\delta_\mathcal{R}(\varphi) = \{\varphi\}$, or
    $$\sum_{\varphi' \in \delta_\mathcal{R}(\varphi)} n_R(\varphi')
    w(\varphi') = n_R(\varphi) w(\varphi)$$ by 
    (\ref{eqn:laststepsoundness}).

    Injecting these relations back into $\NEX{R}{R'}$ we get
    \begin{eqnarray*}
    \NEX{R}{R'} &=& \sum_{\varphi \in \pathsfree{R}} n_R(\varphi)
    \sum_{\varphi' \in 
    \delta_\mathcal{R}(\varphi')} n_\varphi(\varphi') w(\varphi')\\
    &=& \sum_{\varphi \in \pathsfree{R}} 
    n_R(\varphi) w(\varphi) = \NEX{R}{R}
    \end{eqnarray*}
\end{proof}

\begin{corollary}
    \label{cor:nexinv}
    $\mathsf{NEX}$ is an invariant of all weak-reductions starting from $R$.
\end{corollary}

\subsection{Adding promotion : MELL}
We give here a way to add the same operators used by the usual GoI
of MELL to express promotion, while keeping our algebra non-trivial.

Consider $\dlstar_+$ as $\dlstar$ with an added generator
$t$, an unary function $!(.)$ and the axioms 
\begin{eqnarray}
    \dsop{\alpha}{\tau} \dop{\alpha}{\tau} &=& 1 \\
    \dsop{\alpha}{\tau} \oc(a) &=& a \dsop{\alpha}{\tau} \\
    \starred{t} \oc(a) &=& \oc \oc (a) \starred{t} \\
    \modif{\rop{\alpha}} & \modif{\dorth} & \modif{\sop{\alpha}}
\end{eqnarray}
for all $\tau \in \{\wn,\oc\}$ and $a \in \dlstar_+$.
These axioms allow to see the original $\lstar$ monoid of the MELL
GoI as a sub-monoid of $\dlstar_+$. To express the usual $d$ operator
we can either fix a particular name or give any name we want to derelictions.

To assert that $\dlstar_+$ is a non-trivial imz we 
extend $\Mdlstar$ by considering generalized tokens of
the form $(M,E_?,E_!,f,B)$ where
\begin{itemize}
    \item $E_?$ and $E_!$ are now stacks of trees of $\bool$;
    \item $B$ is a stack of pair of such trees.
\end{itemize}
Note that this does not change the validity of the previous definition
of operation in $\Mdlstar$.

We define the following operations :
$$
    (M,\sigma_? \bullet E_?, \sigma_! \bullet E_!,f,B) 
    \operation{\mathfrak{b}} (M, E_?, E_!, 
    f, (\sigma_?,\sigma_!)\bullet B)
$$
$$
    (M, \sigma_? \bullet E_?, \sigma_! \bullet E_!, f, 
    (\tau_?,\tau_!)\bullet B)
    \qquad \qquad
    \qquad \qquad
$$
$$
    \quad \operation{\mathfrak{t}} 
    (M, (\sigma_? \bullet \tau_?) \bullet E_?,
    (\sigma_! \bullet \tau_!) \bullet E_!, f, B)
$$
We can prove that the equations of $\dlstar_+$ are satisfied by this
extension with $!(a) = \starred{\mathfrak{b}} a \mathfrak{b}$.
Note that the self-dual definition of these operations hints at 
a self-dual definition of promotion.

\subsection{MALL}
A simple check of the context semantics used in \cite{MairsonRival02}
to study sharing graphs, which can be seen as another presentation
of the GoI, for MALL allows us to identify it with the 
fragment of $\Mdlstar$ generated by
$\{\Mpop, \Mqop, \Muop{\alpha}, \Mvop{\alpha}\}$. It allows us to assert that
our equational theory is well-suited for the expression of the GoI
of MALL. We will not go into further details as it would require
the introduction of MALL and its GoI. Nevertheless the study of
$\dlstar$ is interesting with respect to additives as it encompasses
this case.

\color{black}

\subsection{Weakenings and shaved execution}
Thanks to the remark prior to theorem \ref{theorem:weightform}
we see that this theorem is still valid for path ending on a 
(co-)weakening and starting either on a free port or another 
(co-)weakening. 
Let $\varphi$ be such a path in the case where it ends on a weakening.
Applying the theorem we get 
$w(\varphi) = \alpha a \starred{b} \starred{\alpha}$.
If $a = \dop{\beta}{\oc} a'$ then by applying backward the 
lemma \ref{lemma:onestepsoundness} we can reconstruct partially
the weak normal form and assert that the weakening will be connected,
with a wire linking their principal ports, to a 
co-dereliction of name $\beta$ in this din.
This path will be destroyed by full
reduction and so is the case of any path starting and ending with
the branch $\alpha$.
We call such branch a \emph{dead} one.

Let 
$$\sigma(\alpha) = 
\left\lbrace \begin{array}{ll}
    0 & \textrm{ when } \alpha = \alpha_0 \alpha' 
    \textrm{ and } \alpha_0 \textrm{ is a dead branch}\\
    1 & \textrm{ otherwise}
\end{array}
\right.
$$
we can refine the execution formula by setting
$$\textsf{SNEX}_R(R') = 
\sum_{
    \scriptstyle 
    \varphi \in \pathsfree{R'} \atop 
    \scriptstyle w(\varphi) = \alpha a \starred{b} \starred{\alpha} \neq 0
}
\sigma(\alpha) n_R(\varphi) w(\varphi)
$$
where the $\textsf{S}$ stands for \emph{shaved}. We have the following 
extension of corollary \ref{cor:nexinv}:
\begin{fact}
    \modif{Let $R$ be a weakly-normal din,}
    $\textsf{SNEX}_R$ is an invariant of all reductions starting from
    $R$.
\end{fact}

\modif{
The restriction to weakly-normal din is to ensure that
no (co-)weakenings can appear by reduction steps.
}
This result is not completely satisfactory
as the argument we have used to express
it is not fitting our initial purpose: checking if a branch is dead requires
a prior computations of the weight of a lot of paths.

\section{Future works}
\modif{
The notion of sub-tree can be made compatible with (\ref{eqn:sumselfdistrib}),
it leads to the notion of \emph{slice} of a relaxed sum-tree.
}
Slices provide a nice framework to express 
properties about non-deterministic computations. 


We can show that all slices of a din have a natural structure of lattice
isomorphic to a sub-lattice of
\modif{
$$\{ \omega \omega^\star,\ \omega \in \dlstarpa\}$$}
In this context complex
assertions about the reduction of a program as a din can be made
into computations of upper and lower bounds.
\modif{This is reminiscent of the usefulness of the lattice
of projectors to study carrier of operators in operator algebra.
Adopting this point of view} could lead to
the introduction of new tools for the study of non-deterministic
computations.

\appendix
\section{Example: a computation of $\mathfrak{S}_2$}
The simple net 
\begin{center}
    \includegraphics[width=5cm]{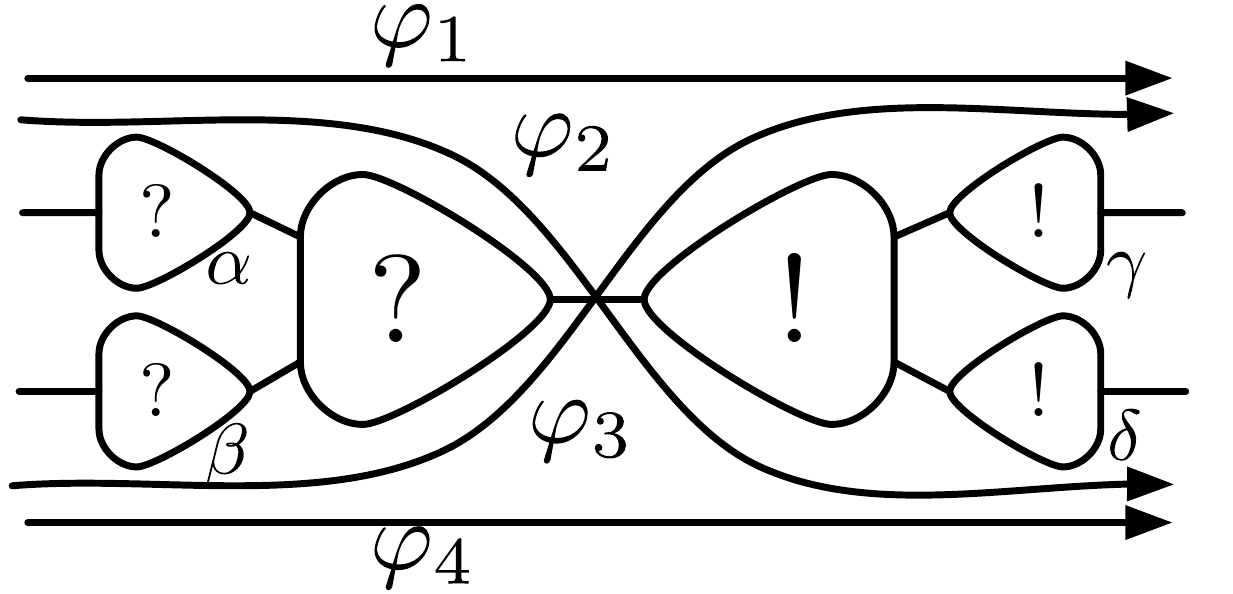}
\end{center}
reduces to the din
\begin{center}
    \includegraphics[width=5cm]{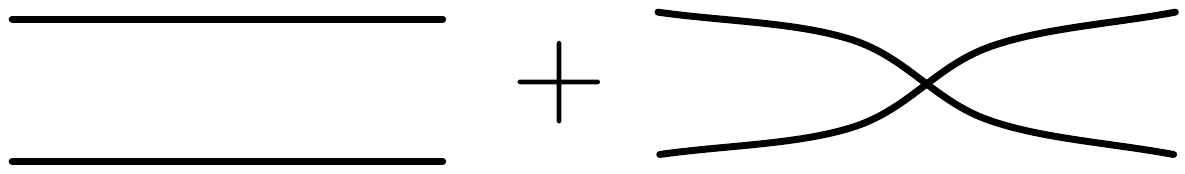}
\end{center}
in many steps of (full) reduction. We can reconstruct
this reduction by using the geometry of interaction.
Computing the weights of the path in the first din we get:
\begin{eqnarray*}
w(\varphi_1) &=& 
\dsop{\oc}{\gamma} \ssop{\oc} \rop{\wn} \dop{\wn}{\alpha} =
\vop{\alpha} \eop{\alpha} \vsop{\alpha}
\uop{\gamma} \eop{\gamma} \usop{\gamma}
\\
w(\varphi_2) &=& 
\dsop{\oc}{\delta} \rsop{\oc} \rop{\wn} \dop{\wn}{\alpha} =
\uop{\alpha} \eop{\alpha} \usop{\alpha}
\uop{\delta} \eop{\delta} \usop{\delta}
\\
w(\varphi_3) &=&
\dsop{\oc}{\gamma} \ssop{\oc} \sop{\wn} \dop{\wn}{\beta} =
\vop{\beta} \eop{\beta} \vsop{\beta}
\vop{\gamma} \eop{\gamma} \vsop{\gamma}
\\
w(\varphi_4) &=& 
\dsop{\oc}{\delta} \rsop{\oc} \sop{\wn} \dop{\wn}{\beta} =
\uop{\beta} \eop{\beta} \usop{\beta}
\vop{\delta} \eop{\delta} \vsop{\delta}
\end{eqnarray*}

We set $\delta(\varphi_i) = \{\varphi'_i\}$. We can reconstruct
from the weight of $\varphi_i$ the branch prefix from which 
$\varphi'_i$ is present in all subsequent leaves:
\begin{center}
    \includegraphics[width=5cm]{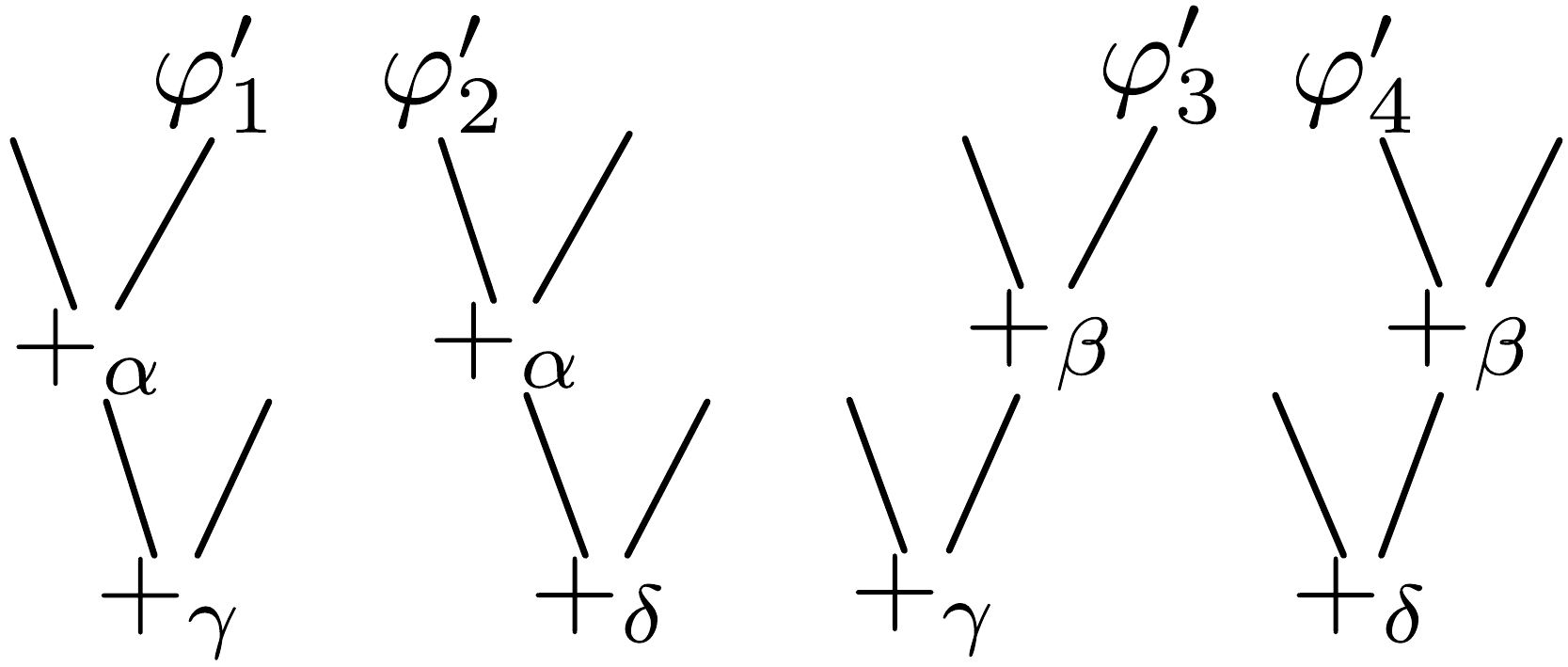}
\end{center}


We can reconstruct the full weak-normal form
and a simple check allows us to assert that any leaf not drawn will
be shaved by full reduction:

\begin{center}
    \includegraphics[width=7cm]{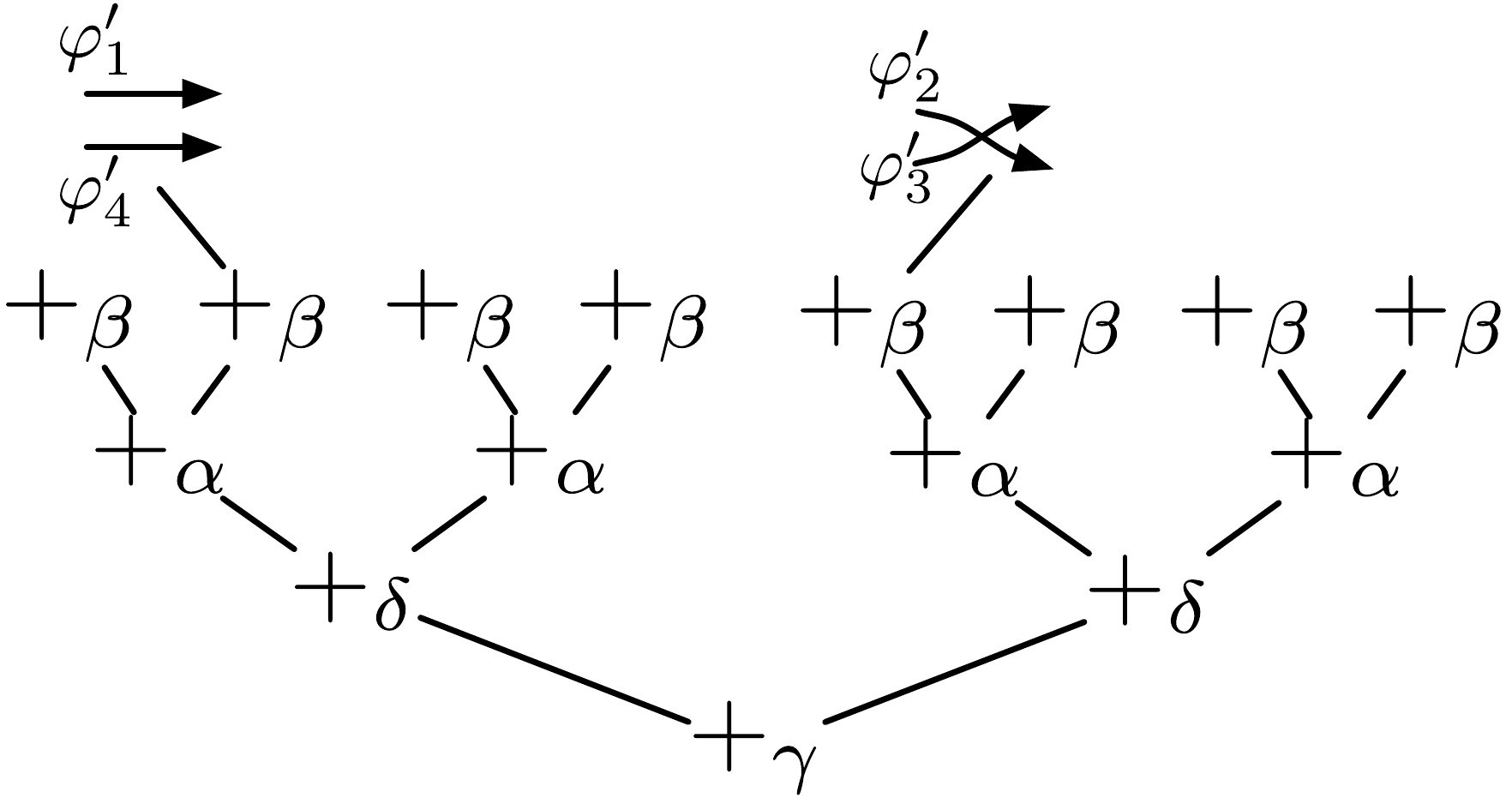}
\end{center}

\color{black}

\bibliographystyle{latex8}
\bibliography{biblio}

\end{document}